\documentclass[preprint]{aastex}

\shorttitle{W51} \shortauthors{Figuer\^edo et al.}

\usepackage[dvips]{color}

\definecolor{darkred}{rgb}{0.6,0.0,0.0}

\begin{document}

\title{The Stellar Content of Obscured Galactic Giant HII Regions. VI: W51A}

\author{E. Figuer\^edo\altaffilmark{1}}
\affil{Department of Physics and Astronomy, The Open University, Milton Keynes,
MK7 6AA, UK}
\email{E.Figueredo@open.ac.uk}

\author{R. D. Blum} \affil{National Optical Astronomy Observatory, 950
North Cherry Aveneue, Tucson, Arizona, 85719} \email{rblum@noao.edu}

\author{A. Damineli\altaffilmark{1}}
\affil{IAG--USP, R. do Mat\~ao 1226, 05508--900, S\~ao Paulo, Brazil}
\email{damineli@astro.iag.usp.br}

\author{P. S. Conti}
\affil{JILA, University of Colorado \\ Campus Box 440, Boulder, CO, 80309}
\email{pconti@jila.colorado.edu}

\and

\author{C. L. Barbosa}
\affil{IP\&D, Universidade do Vale do Para\'{\i}ba, Av. Shihima Hifumi 2911, S\~ao
Jos\'e dos Campos 12244-000, SP, Brazil}
\email{cassio@univap.br}

\altaffiltext{1}{Visiting Astronomer, Cerro Tololo Inter--American Observatory,
National Optical Astronomy Observatories, which is operated by Associated
Universities for Research in Astronomy, Inc., under cooperative agreement with
the National Science Foundation.}

\begin{abstract}

We present $K$--band spectra of newly born OB stars in the obscured
Galactic giant H II region W51A and $\approx 0.8''$ angular resolution
images in the $J$, $H$ and $K_S$--bands. Four objects have been
spectroscopically classified as O--type stars. The mean spectroscopic
parallax of the four stars gives a distance of 2.0 $\pm$ 0.3 kpc
(error in the mean), significantly smaller than the radio
recombination line kinematic value of 5.5 kpc or the values derived
from maser propermotion observations (6--8 kpc). The number of Lyman
continuum photons from the contribution of all massive stars (NLyc
$\approx$ 1.5 $\times 10^{50}$ s$^{-1}$) is in good agreement with that
inferred from radio recombination lines (NLyc = 1.3 $\times 10^{50}$
s$^{-1}$) after accounting for the smaller distance derived here.

We present analysis of archival high angular resolution images (NAOS
CONICA at VLT and T-ReCS at Gemini) of the compact region W51
IRS2. The $K_S$--band images resolve the infrared source IRS~2
indicating that it is a very young compact HII region. 
Sources IRS2E was resolved into compact cluster (within 
660 AU of projected distance) of 3 objects, but one of them is just 
bright extended emission. W51d1 and W51d2 were identified with compact 
clusters of 3 objects (maybe 4 in the case of W51d1) each one.
Although IRS~2E is
the brightest source in the $K$--band and at 12.6 \micron, it is not
clearly associated with a radio continuum source. Our spectrum of
IRS~2E shows, similar to previous work, strong emission in Br$\gamma$
and HeI, as well as three forbidden emission lines of FeIII and
emission lines of molecular hydrogen (H$_2$) marking it as a massive
young stellar object.

\end{abstract}

\keywords{HII regions --- infrared: stars --- stars: early-type --- stars:
fundamental parameters --- stars: formation}

\section{Introduction}

The study of Giant HII regions (GHII -- here taken to emit at least
$10^{50}$~LyC photons s$^{-1}$, or $\approx$~10 $\times$ Orion) in the
near--infrared can address important astrophysical issues. The
exploration of the stellar content of obscured Galactic GHII regions
has been studied recently by several groups: \citet{han97},
\citet{blum99,blum00,blum01}, \citet{fig02, fig05}, \citet{okum00}
and \citet{bik05,bik06}. These observations revealed massive star
clusters at the centers of the HII regions which had been previously
discovered and studied only at much longer radio wavelengths.

The first step needed to trace the spiral structure of our Galaxy
based on the spatial distribution of giant HII (GHII) regions is to
derive spectroscopic parallaxes of OB stars in each embedded
cluster. This is a well tested method in the optical window. We work
in the near-infrared to overcome the obscuration at the optical
wavelengths (A$_K$/A$_V$ $\approx$~ 0.1) and use the recently
developed spectral classification scheme in the near infrared
\citep{han96} to identify and classify the specific types of massive
stars present. The uncertainty in the distance estimate is dominated
by the scatter in the intrinsic $M_V$ of O--type stars \citep[$\pm$
0.67 mag - ][]{vac96}, leading to an uncertainty of 30\% for a single
O--type star. The uncertainty is diminished as spectra are
accumulated, and can reach $<$10\% if enough stars are observed. If
B--type stars are included the individual accuracy is better as their
$M_K$ are more constrained given their longer main sequence lifetimes.

Our procedure is independent of the traditional kinematic distance
method based on radial velocity measurements of radio recombination
lines and adoption of a Galactic rotation model. Almost all of the
currently determined distances of Galactic GHII are based on this
``kinematic'' procedure which contains an ambiguity for directions
inward of the solar circle and uncertainties introduced by random
velocity components in the gas. In our previous work we were able to
break the distance ambiguity of W31 \citep{blum01}, placing it on the
near side of the GC at 3.4 $\pm$ 0.5 kpc from the Sun. In the case of
W43 \citep{blum99} and W42 \citep{blum00} we also derived smaller
distances than those inferred from the pure rotational model; in
G333.1-0.4 \citep{fig05}, we also derive a smaller distance than given
by the kinematic method (2.6 $\pm$ 0.4 kpc). Similar results have been
found by other investigators using the same technique \citep{han97,
bik04} for a number of other GHII. A recent H$_2$O maser study of the
star forming HII region W3 finds a similar result based on maser
parallax \citep{xu06}; i.e., W3 is closer than the kinematic distance
and consistent with OB star absolute magnitudes in nearby associations
lending credence to our results.

W51 is a complex of giant HII regions that was first detected by its
free--free emission in the radio continuum \citep{wes58}. The radio
observations show that W51 is, in fact, composed of a group of
individual sources spread in an area of one degree in the Galactic
plane \citep{bie75}. The most luminous region in the W51 complex is
G49.5-0.4 (R.A.$=$ 19h23m42.02 and Dec. $=$ $+$14d30m33.56s J2000),
which is divided into eight smaller radio sources: W51A to W51H,
ordered by R.A. W51A, one of the most luminous regions in the W51
complex, is located at a kinematic distance 5.5 kpc (near 
distance), adopting the value given by \citet{russ03} who used the 1985 IAU 
recommended values
for the Sun to Galactic Center (GC) distance and rotation speed of the
Milky Way, which are R$_0$ = 8.5 kpc and $\theta_0 = 220$ km s$^{-1}$.
Although it is one of the most luminous star-forming regions in the
Galaxy, this source is completely obscured in the optical wavelengths
due to the large extinction at optical wavelengths along the line of
sight.

\citet{gol94} have imaged an area of 90" $\times$ 90" in the direction
of W51A, using an angular resolution of $\approx 0.4''$ in the near
infrared. In addition to the radio sources that had been previously
discovered, IRS1 and IRS2, they identified an third bright source --
IRS3. They also argued that IRS2 is in fact a small cluster with at
least twelve point sources in $K$--band. They have compared the images
from a narrow band filter, centered in $Br\gamma$, with $K$--band
images, and they have concluded that the extended inter--cluster
emission is thermal from ionized hydrogen and neutral helium. The
surface brightness of the extended emission in the radio and
near--infrared led to a extinction of $A_{K_{S}} = 2.6 \pm 0.3$ mag for
W51A, with areas still more obscured. We used this work for the
identification of candidates for spectroscopy in order to determine
the distance of W51A by the spectroscopic parallax method.

\citet{okum00} studied the properties of the stellar content of  W51A
in order to determine the stellar masses, ages, spatial distribution
and IMF. They used $J$, $H$ and $K$ imaging on an area of 15' $\times$
15' covering the G49.5-0.4 complex. These authors derived the
interstellar extinction in the W51A direction by  using the extinction
in the visual band determined in a control field 20$'$ from the source
IRS2 along the Galactic plane. The histogram of the number of
stars in each extinction bin shows two peaks at $A_{V} = 7$ and $A_{V} =
19$ mag. \citet{okum00} explained this effect as the presence of the
Sagittarius and Perseus spiral arms. By considering the distances of
five kpc and 13 kpc for the two arms \citep{gg76}, respectively,
\citet{okum00} derived an average interstellar differential extinction
of 1.2 and 1.5 mag kpc$^{-1}$ in the visual band. \citet{okum00} also
found that the IMF slope in W51A agrees with the IMF slope in the
control field in the range of masses between 10 and 30  M$_\odot$. On
the other hand, they found an excess of stars more massive than 30
$M_{\odot}$.

In the present work, we are interested primarily in constraining the
distance to W51A using spectroscopic parallaxes of newly born OB
stars. In fact, with 8.71 $\times$ 10$^{50}$ Lyman continuum photons
per second \citep[=NLyc --][]{cc04}, this would be equivalent to
$\approx$ 20 O3V-type stars or $\approx$ 200 O7V-type stars
\citep{mar05}. We present an investigation of the stellar content of
W51A through $JHK_S$ imaging and $K$--band spectroscopy (described in
\S2). In \S3 we consider the photometry results, and in \S4 we analyze
the spectra. We determine the distance to W51A in \S5 and discuss the
results in \S6.

\section{Observations and Data Reduction}

$K$--band ($\lambda$ $\approx$ 2.2 $\mu$m, $\Delta$$\lambda$ $\approx$
0.33 $\mu$m) spectra of five objects in W51A were obtained on the
nights of 2004 May 08 and 09 and 2004 June 01 using
NIRI\footnote{Gemini's Near InfraRed Imager and spectrograph was built
  by the University of Hawaii's Institute for Astronomy.  See
  \citet{hod03} for more details about the instrument.}  on
Gemini-North in queue-scheduled mode (GN-2004A-Q13). NIRI in $f$/6
mode delivers a plate scale of 0.12$''$ pix$^{-1}$. The $f$/6 camera
and a four--pixel long slit ($0.48'' \times 110''$) were used. The $K$
grism produces $K$--band spectra with a resolution of $R$ $\approx$
780 for NIRI and the linear dispersion is $\lambda/pix$ $\approx$ 7.08
\AA /pixel.

$J$ ($\lambda$ $\approx$ 1.25 $\mu$m, $\Delta$$\lambda$ $\approx$ 0.2
$\mu$m), $H$ ($\lambda$ $\approx$ 1.64 $\mu$m, $\Delta\lambda$
$\approx$ 0.3 $\mu$m) and $K_S$ ($\lambda$ $\approx$ 2.15 $\mu$m,
$\Delta$$\lambda$ $\approx$ 0.3 $\mu$m) images of W51A were obtained
on the night of 2005 May 07 using the facility infrared camera
ISPI\footnote{ISPI (Infrared Side Port Imager -- ``eye-spy''). More
details about the instrument are available at the CTIO webpage.} at
the CTIO Blanco 4-m telescope (see \citet{pro03}).  ISPI in F/6 mode
delivers a plate scale of 0.3$''$ pix$^{-1}$ and a field of view of
$10.25' \times 10.25'$.

All basic data reduction was accomplished using IRAF\footnote{IRAF is
distributed by the National Optical Astronomy Observatories.}. Each
image was flat--fielded using dome flats and then sky subtracted using
a median--combined image of independent sky frames obtained 10$'$
North and 10$'$ East from W51A for direct imaging and three dithered
frames 1$'$--2$'$ far from the target for spectroscopy. One
dimensional spectra were obtained by extracting and summing the flux
in $\pm$ 4 pixel apertures. The extractions include local background
subtraction from apertures $\approx$ 1$''$ on either side of the
object.

\subsection{Photometry}

The images of W51A were obtained in non--photometric (cirrus)
conditions.  Total exposure times were 15, 3.4 and 2.4 minutes at $J$,
$H$ and $K_S$, respectively. The individual $J$, $H$ and $K_S$ frames
were shifted and combined. These combined frames have point sources
with FWHM of $\approx$ 0.92$"$, 0.91$"$ and 0.78$"$ at $J$, $H$ and
$K_S$, respectively.

\begin{figure}[!ht] \begin{center}
\includegraphics[totalheight=13.0cm,angle=0]{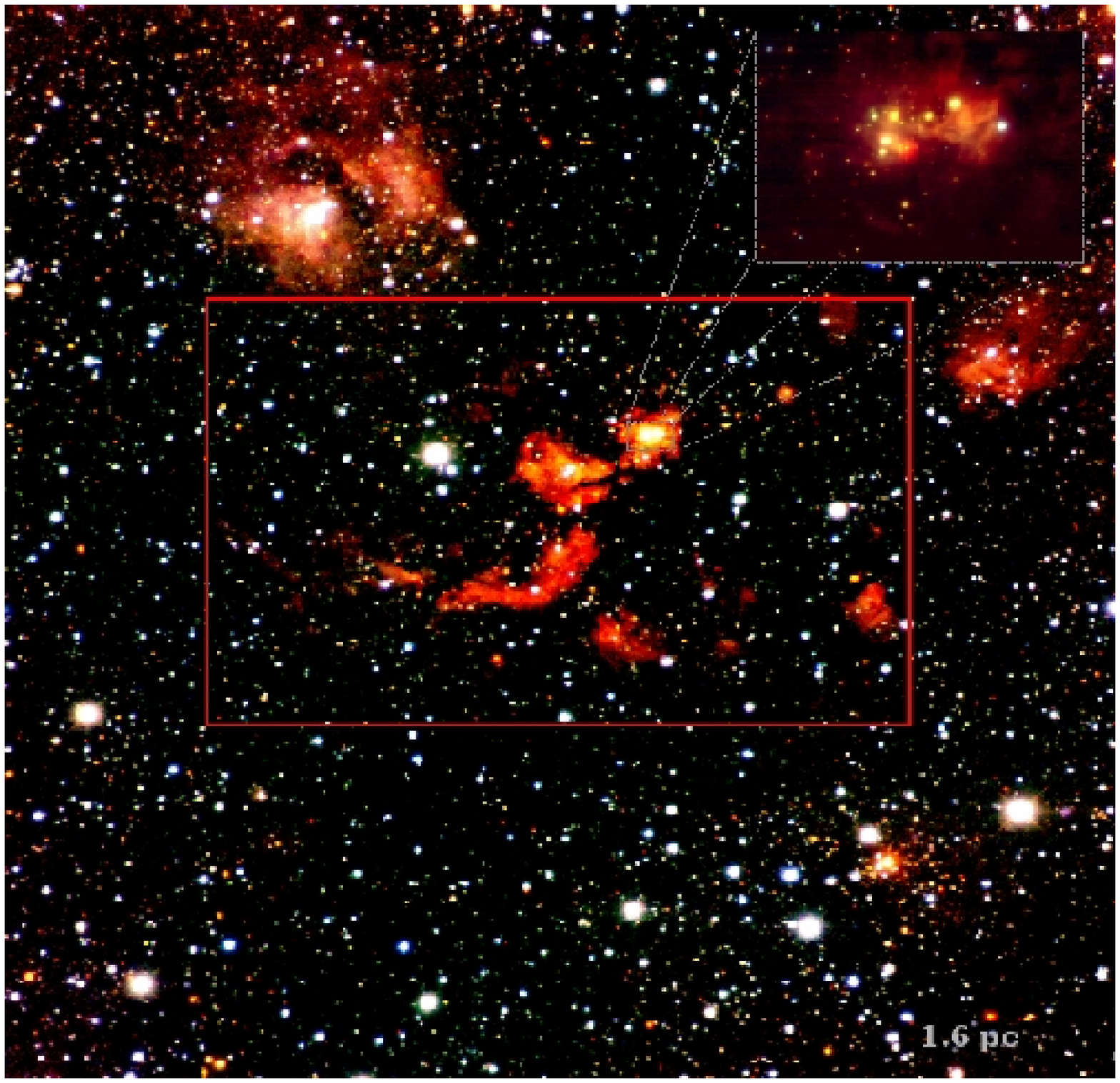}
\caption{False color image of W51A: $J$ is blue, $H$ is green and $K_S$ is red.
The red rectangle indicates the region where DoPHOT photometry  was performed.
The coordinates of the center of the image are RA (2000) =
19h23m42.02s and Dec. = $+$14d30m33.56s and the size of the image is 10$'$
$\times$ 10$'$ (plate scale = 0.3"/pixel). North is up and East to the
left. The zoomed box (top right) shows a VLT false color image of W51 IRS2 (see
\S5.1 and also Figures~\ref{nacoh} and~\ref{irs2-mir}). 
\label{color}} \end{center} 
\end{figure}

Figure~\ref{color} shows a false color image, made by combining the
three near infrared images and adopting the colors blue, green and
red, for $J$, $H$, and $K_S$, respectively. This Figure shows two
regions with strong Br$\gamma$ emission probably associated with
embedded star clusters seen in the center \citep[W51A -- region 3
from][]{okum00} and in the NE of the image \citep[region 2 from
][]{okum00}. The bluest stars are likely foreground objects, and the
reddest stars are probably $K_S$--band excess emission objects,
indicating the presence of hot dust associated with recently formed
stars in the cluster. Background objects seen through a high column of
interstellar dust would also appear red (but also fainter). The
rectangle in Figure~\ref{color}, indicates the region, including W51A,
where DoPHOT~\citep{sch93} photometry was performed on the combined
images. DoPHOT detected a total of 2419 objects inside this rectangle
in the $K_S$--band image. Of those 2419 objects, 1338 were also
detected in $H$--band and 855 in all three filters. All objects
detected in $J$ or $H$--bands that were not picked up in $K$ were
excluded from our catalog because their errors were bigger than the
cutoff limit (see below). Figure~\ref{finding} shows a finding chart
using the $K_S$--band image corresponding to the rectangular area of
Figure~\ref{color}. The numbers in the figure correspond to the brighter 
sources detected by DoPHOT.

Standard stars were not observed but flux calibration was accomplished
using stars in the same field, available in the 2MASS\footnote{2MASS:
Two Micron All Sky Survey \citet{skr06} - available at
http://www.ipac.caltech.edu/2mass/releases/allsky/} All-Sky Point
Source Catalog (PSC). Twenty uncrowded stars on the W51A images were
used for this purpose. The DoPHOT magnitudes were compared to the
2MASS apparent magnitudes in order to define a zero point to the
photometric calibration. The rms scatter in the zero point (difference
between DoPHOT and 2MASS) in $J$, $H$, and $K_S$ was 0.032, 0.051, and
0.082, respectively. The error in the mean for the twenty stars is
then 0.007, 0.011, and 0.019 in $J$, $H$, and $K_S$, respectively.
Uncertainties for the $J$, $H$ and $K_S$ magnitudes in W51A include
the formal DoPHOT error added in quadrature to the error in the mean
of the zero point offset. The sum in quadrature of both uncertainties
results in typical bright star uncertainties of $\pm$ 0.019, $\pm$0.028
and $\pm$0.038 mag in $J$, $H$ and $K_S$, respectively. The derived
magnitudes and colors of the stars labeled in Figure~\ref{finding} are 
given in the Table~\ref{stars}.

The completeness of the DoPHOT detections was explored through
artificial star experiments in the rectangular region of Figure 1 (see
the discussion in Figueredo et al. 2005 for a description of the
technique).  The completeness is better than 90\% for $J$ and $H <
16.5$ and for $K_S$ $<$ 16. In other words, the detection of faint
sources is more complete in $J$ and $H$--bands than in $K_S$. This is
likely because we are probably accessing mainly the (dense) foreground
population and most of the fainter cluster members are not detected at
this angular resolution (i.e., the true completeness limits in the
{\it clustered regions of the image} are likely much brighter). 
The strong nebulosity will also hamper the detection of faint objects in the 
cluster region. Our primary goal in this paper is to derive the distance to W51A, and hence we only rely on the much brighter 
massive members of the cluster where crowding is not a significant factor. 
Indeed, the spectroscopic targets described below are in less crowded regions 
of W51A.

The mean of the differences between the input and output $H$ and $K_S$ 
magnitudes of the artificial stars experiment, considering only the range 
of magnitudes where the completeness is better than 90 \%, are $\approx 0.07$ 
and $0.11$, respectively. The results of this experiment are similar to the 
ones showed in \citet{fig05}. The magnitudes errors determined by this method 
are much more realistic than the formal errors defined by DoPHOT.  We adopted
an arbitrary cutoff of 0.2 mag (stars with larger errors were excluded
from further analysis).

\subsection{Spectroscopy}

In order to avoid spectral degradation by bad pixels, the total
exposure time needed to obtain $S/N \sim 100$ was split and individual
spectra were obtained at seven different positions along the
slit. After each exposure, the telescope was offset by three arcsec
along the slit. In two cases, two or three stars could fit in the same
slit, therefore the exposure time was adjusted to the fainter star. We
needed blank sky spectra in order to subtract the sky emission of each
target spectrum that is contaminated by nebular emission. The sky
background were obtained by median combination of three dithered
exposures in a blank field one or two arcminutes from each target
(total = 10 sky positions).  Following this procedure, the $K$--band
spectra of five of the brightest stars in the W51A cluster were
obtained: RS6 (in this paper: \#50 - see Figure~\ref{finding}), RS7
(\#44) and IRS2East (IRS2E) from \citet{gol94} and \#2519 (\#61) and
\#2670 (\#57) from \citet{okum00}.

One dimensional spectra were obtained by extracting and summing the
flux in $\pm$ eight pixel apertures (0.96'') centered on the target in
each dithered position. The extractions include local background
subtraction from apertures $\approx$ 1$''$ on either side of the
object. Wavelength calibration was accomplished by using the spectrum
of an Argon lamp obtained each night. The error of this calibration is
$< 0.8$ \AA \ ($\sim 10\%$ of a pixel). In order to remove telluric
features in the spectra of W51A stars, the A0V stars HIP92177 and
HIP98640 were observed using an identical procedure, just after or
before those obtained for the target.  At the time of the observation,
both A0V stars have an air mass very close to the mean air mass
attained by the targets in W51 during the observation period (Delta
($\Delta$) $\approx 0.02$ airmass). The Br$\gamma$ photospheric feature
was removed from the average A0V--type star spectrum by fitting and
subtracting a Lorentz profile. Since the Br$\gamma$ region is
relatively free from strong telluric features, this procedure is
enough to obtain the template for telluric lines. The wavelength
calibrated spectra were then divided by the continuum of the A0V--type
star to remove telluric absorption.

The $K$--band classification scheme for OB stars is based on the
(relatively weak) lines of CIV, HeI, NIII and HeII. The spectra of
young stars in HII regions are often contaminated by the 2.058 $\mu$m
HeI and Br$\gamma$ nebular lines but these features are not used for
classifying O--type stars \citep{han96}.

After the observations were obtained, it was found that \citet{ma96}
have identified the ``A0V'' telluric star HIP92177 (KOAql) as a binary system,
and the secondary star is cooler than a G0--type star.  Fortunately,
the spectra of G--type stars don't show any strong features near the
most important lines used to identify the O--type stars that are
located between 2.05 $\mu$m and 2.2 $\mu$m. In fact, the Calcium, Iron
and Sodium lines, together with the CO absorption bands, common in
cool stars \citep{lan92}, are located at wavelengths longer than
2.2 $\mu$m. In this work we thus consider the star HIP92177 as a telluric
star, ignoring all lines longer than 2.2 $\mu$m wavelength.

\section{Results of the Photometry}

The ISPI $J$, $H$ and $K_S$--band images reveal W51A as an embedded
star cluster readily seen in the center of Figure~\ref{finding}.
Due to the strong nebulosity it is hard to detect the faintest stars in this
region. On the other hand, a comparison between the density of stars in the 
center of this figure with a sky image close to W51A, allow us to verify 
that the stellar density is higher in the center than the other areas of the
image.

\citet{okum00} found W51A is younger than $0.8$ million
years. This is in agreement with the photometric characteristics of
this region, such as the high local reddening indicating that the
massive stars have not completely shed their birth cocoons, and it is
likely they are on the zero age main sequence (ZAMS). The cluster
reddening can be approximated as $A_{K_S} \approx 1.7 \times (H -
K_S)$ since the intrinsic colors of OB stars are nearly zero
\citep{koo83}; see \citet{ccm89,mat90}.

The $H - K_S$ versus $K_S$ color magnitude diagram (CMD) is displayed
in Figure~\ref{cmd}. The labels in all plots refer to the same stars
as in Figure~\ref{finding}. Both vertical lines indicate the position
of the theoretical ZAMS shifted to 2.0 kpc (\S6) and with interstellar
reddening $A_{K_S} = 0.65~mag$ (dashed line) and an additional ``intra
cluster'' reddening component of $A_{K_S} = 3.27~mag$ ($A_{K_S total}
= 3.92~mag$ -- {\it solid} line). The strong nebulosity in the cluster 
region, together with the crowding, can explain the gap at $H-K_{S} 
\sim 2$ that appear in Figure~\ref{cmd} between $K=13$ and 
$K=15$ magnitudes. This gap is probably generated by the incompleteness 
as described in \S2.1. 

The $J - H$ versus $H - K_S$ color--color plot is displayed in
Figure~\ref{ccd}. In that diagram the solid lines, from top to bottom,
indicate interstellar reddening for main sequence M--type
\citep{fro78}, O--type \citep{koo83} and T~Tauri \citep{mey97} stars
(dashed line). The solid vertical line between the main sequence
M--type and O--type lines indicates the position of ZAMS. Asterisks
indicate $A{_{K_S}}$ = 0, 1, 2 and 3 extinction values (in magnitude).
Dots are objects detected in all three filters. The error bars in both
figures refer to the final errors in the magnitudes and colors. The
small number of objects between both solid lines in Figure~\ref{ccd}
means that most of stars have not been detected in the $J$--band
images due the high extinction at this wavelength ($A_J \sim
10.5~mag$). Objects found to the right of the O--type star 
reddening line in the
CMD of Figure~\ref{ccd}, shown as a solid diagonal line, have colors
deviating from pure interstellar reddening. This is frequently seen in
young star clusters and is explained by hot dust in the immediate
circumstellar environment. 

For cases where the object has an apparent $J-H$ less than an
unreddened O star, the explanation is not clear. Most of cases, but not all, 
those objects are faint stars. Such objects may
actually be blends of two nearby stars of different colors which are
inadvertently matched when merging the photometry from different
filters. This can happen when a red cluster star is projected near to
a bluer foreground star. 

\subsection{Reddening and Excess Emission}

Two concentrations of objects are seen in the CMD around the two
vertical lines. The first one appears around $H - K_S \approx 0.4$,
which corresponds to an extinction of $A_{K_{S}} = 0.65~mag$ $(A_{V}
\approx 6.5~mag)$ using the interstellar reddening curve of
\citet{mat90}. This concentration represents foreground stars
(see also Figure~\ref{ccd}). The foreground extinction can be estimated as 
$A_{K_{S}} \approx
0.3~mag/kpc$ when using the distance of 2.0 kpc to W51A. \citet{jk94} found 
smaller values, $A_{V} \approx 1.8-2.5~mag/kpc$ or $A_{K} \approx 0.18-0.25~mag/kpc$, for 
a field close to the galactic plane. On the other hand, \citet{okum00} 
defined as foreground sources in the direction of W51A cluster, all objects 
with $A_{K} \le 0.9~mag/kpc$. The second and smaller concentration of 
objects appears around $H - K_S \sim 2.0$ 
or $A_{K_{S}} = 3.27~mag$, indicating the average color of cluster members. A
number of stars display much redder colors, especially brighter
$K$--band sources. These objects are located at $H - K_S > 2.0$ and
$A_{K_{S}}> 3.2$.

Determining the local reddening of a young cluster such as W51A is not
simple, as the concentration of stars does not appear as a clear main
sequence, and the scatter about the mean value of $A_{K_{S}}$ is high. In
fact, the differential ``intra-cluster'' reddening and the inclusion
of very young sources produces the variation of the points on the CMD
for $H - K_S > 0.4$. The local reddening can be determined through the
colors of the O--type stars we have classified (see \S4.1). The mean
color of the four stars spectroscopically observed are $(H-K_S) =
2.25$, which gives $A_{K_S} = 3.92~mag$. The resulting local reddening
in W51A direction, after subtracting the interstellar component, is
then $A_{K_S} \approx 3.27~mag$ ($A_{V} \approx 33~mag$).

In order to place the ZAMS in the CMD, as Figure~\ref{cmd} shows, the
reddening and the distance of W51A must be derived. The
procedure used to transform bolometric magnitudes ($M_{bol}$) and effective
temperatures ($T_{eff}$) into colors ($H-K$) and absolute $K$ magnitudes has
been given by \citet{blum00}. No transformation has been derived between ISPI and the system used by 
\citet{blum00} (CIT/CTIO). The difference between both systems are negligible 
in comparison with the final uncertainties in the photometry. Figure~\ref{cmd} shows the ZAMS as a
vertical dashed line, shifted to $D = 2.0$ kpc and reddened by  $A_{K_S} =
0.65$ mag due to the interstellar component. When adding the average local
reddening ($A_{K_S} = 3.27$ mag), the ZAMS line is displaced to the right and
downward, as showed by the solid line. A 5.5 kpc distance for
W51A would move this ZAMS downward in the CMD by two magnitudes and the
{\it observed} magnitudes for the spectroscopic targets would be
inconsistent with a ZAMS luminosity class. 

Besides the strong local reddening in the direction of W51A, the
color--color diagram and CMD show several objects whose colors deviate
from the pure interstellar reddening. These objects are located to the
right of the O star reddening line in the CMD of Figure~\ref{cmd}. We
can estimate a lower limit to the excess emission in the $K$--band by
supposing that the excesses at $J$ and $H$ are negligible, and that
the intrinsic colors of the embedded stars correspond to normal OB
stars. Indeed, we can follow the same procedure as \citet{fig02} by
assuming that our sample of stars is composed of young objects (not
contaminated by foreground or background stars) which have intrinsic
colors in the range $(H - K)_\circ = 0.0 \pm 0.06$ mag
\citep{koo83}. In \citet{fig05} we adopted for all objects the
intrinsic colors of a B2~V star and from the difference between the
observed $J - H$ and the adopted B2~V intrinsic $J-H$ color, we
obtained the $J$--band extinction by using the adopted extinction law
\citep{ccm89, mat90}. The results for the same analysis in W51A are
displayed in Figure~\ref{excess}. 

In Figure~\ref{excess} the solid line indicates $K_{exc}=0$. Connected
solid diamonds refer to the average value of the $K$ excess in 1
magnitude bins.  Dashed lines indicate 1, 2 and 3 $\sigma$ from the
average. Bright objects with very large excess (very positive values)
in Figure~\ref{excess} cannot be explained by errors in the
dereddening procedure. The excess should be due to accretion disks
and/or hot dusty envelopes around the less massive objects in the
cluster. A few objects are well above the 3 $\sigma$ scatter for
otherwise normal stars in Figure~\ref{excess}.

As with in the case of G333.1-0.4, most of the stars in W51A
(Figure~\ref{excess}) have a modest negative excess, about $0.2$
magnitude.  This negative excess has been explained by \citet{fig05}
as a consequence of our assumption that all stars have the intrinsic
color of a B2~V type star; in reality, most of the stars will have
slightly positive intrinsic color. Our goal is to identify stars with
a significant excess which would cause them to lie to the right of
reddening line in Figure~\ref{ccd}, so this modest negative excess for
``normal'' stars, or stars with a small excess is not important for
our purposes. In the following sections we have only corrected the
$K$--excess for stars with positive excess as determined here, for all
others we impose zero excess emission.

It is important to note that the excess emission as determined in this
section is only a lower limit, since we assumed the excess was primarily in
the $K$--band (but according to Figure~\ref{excess}, there are not many stars
with large excess for the higher masses). \citet{hsvk92} have computed disk
reprocessing models which show the excess in $J$ and $H$ can also be large for
disks which reprocess the central star radiation. The $K_S$--band excess
emission determined for IRS2E, about two magnitudes, is in agreement with the
values found by \citet{hil00} for young stars in the Orion cluster. Objects on
the top of Figure~\ref{excess} show excess emission still higher and it is
an indication that probably such peculiar objects are, in fact, massive YSOs or
ultra-compact HII (UHII) regions, as should be the case of IRS2E. \citet{gol94}
have found, besides IRS2, a pair of luminous sources in W51A - IRS1 and
IRS3. The coordinates of IRS1 match those of \#403 in figure~\ref{finding}, 
and IRS3 with \#96 and \#152, probably IRS3E and IRS3W
respectively. Only \#152 has been detected in $H$ and $K$--band.
Lower mass protostars also can show excess emission in $K$--band.

As we can see in Figures~\ref{cmd}, \ref{ccd} and \ref{excess}, some objects in
W51A are very bright, display very red colors and at the same time have
excess emission well above the 3 $\sigma$ scatter for the average of normal
stars. This is the case for objects such as IRS2W. 

Given such evidence for circumstellar disks, we attempt to estimate by 
iteration the excess emission using \citet{hsvk92} models  for reprocessing 
disks, such as was done in the case of G333.1-0.4 \citep{fig05}, to get a 
sense for how bigger the excess might be compared to that derived  by 
assuming all the excess is in the $K_S$--band. This was done for 
the star we have detected in all three bands, IRS2W. The results corresponds 
to an excess emission of $K_{exc} = 3.58~mag$ from a face-on reprocessing disk with
a central source of $15 ~M_{\odot}$ corresponding to the spectral type B0.5.
The resultant reddening for this star was determined by using \citet{mat90}
approximation as  $A_{K_S} \approx 1.7 \times ((H-K_S) + 0.04 - 0.5)$. The resultant
extinction for IRS2W, after correcting for the excess as determined above, 
is $A_{K_S} = 4.25~mag$. 

\section{Results: Analysis of Spectra \label{espectro}}

The Gemini North/NIRI spectra of five objects in W51 sources are shown
in Figures~\ref{spec1} and \ref{spec2}. All sources have been divided
by a low--order fit to the continuum after correction for telluric
absorption. The signal--to--noise ratio is S/N $\approx$ 100. These
spectra have been background subtracted with nearby ($\approx 1''.0$)
apertures, though non--uniform extended emission could affect the
resulting He~I and Br$\gamma$ absorption apparently present in the
stars themselves.

In order to identify the spectral type of the stars we compared all
spectra with the $K$--band spectroscopic standards presented by
\citet{han96} and \citet{bik05}. The features of greatest importance
for classification are (vacuum wavelengths) the CIV triplet at 2.0705
$\mu$m, 2.0769 $\mu$m and 2.0842 $\mu$m (emission), the NIII complex
at 2.116 $\mu$m (emission), and HeII at 2.1891 $\mu$m
(absorption). Stars cooler than spectral type O7 show HeI (absorption)
to the blue side of the NIII complex, at 2.1126 $\mu$m (which can be
confused with nebular over--subtraction).

The NIII complex has been identified by \citet{han96} in early-O stars
as being due to NIII(8--7). The fact that this transition is a
multiplet system explains the broad profile of the feature, centered
at 2.116 $\mu$m.  It shows constant width through moderate variations
in temperature and large variations in luminosity. The CIV triplet
(2.071 $\mu$m, 2.077 $\mu$m and 2.084 $\mu$m) is typically weak and
seen only in very high signal--to--noise spectra \citep{han96}. The
present classification system laid out by \citet{han96} does not have
strong luminosity--class indicators. Still, without nebular
contamination (which is not the case presently), the HeI (2.0581
$\mu$m) and Br$\gamma$ (2.1661 $\mu$m) features can be used to
approximately distinguish between dwarfs plus giants on the one hand,
and supergiants on the other \citep{han05}. Generally strong
absorption in Br$\gamma$ is expected for dwarfs and giant stars and
weak absorption or emission for supergiants.

\subsection{O--type Stars}

Figure~\ref{spec1} shows four spectra that may be compared with the
$K$--band spectroscopic standards presented by \citet{han96}. The
signal--to--noise ratio is $S/R \sim 100-120$, high enough to separate
and identify the CIV triplet and NIII complex lines. The telluric
absorption lines have been removed from the \#44, \#50 and \#61
original spectra by using the HIP92177 spectrum. The spectrum of star
\#57 has been divided by HIP98640. The measured equivalent widths for
the main lines for classification of the stars in Figure~\ref{spec1}
are shown in the Table~\ref{EWS}. The classification resulted in the
spectral types indicated in the Figure.

The CIV triplet at 2.069-2.083 $\mu$m (emission) and the NIII complex
at 2.116 $\mu$m (emission) identify the four sources in
Figure~\ref{spec1} as early to mid O-type stars. The relatively strong
NIII and HeII lines that appear in the spectrum of \#57 indicate that
this star is hotter than O6. On the other hand, the absence of CIV at
2.071 $\mu$m places \#57 in the kO3--O4 subclass \citep{han96}. The
CIV line appears very weak and blended with the third line of the
triplet. The CIV line in the spectrum of the source \#44 indicates
that this star is cooler than a typical O4 and hotter than an O7-type
star. The NIII line is still stronger than in the previous star,
placing \#44 in the kO5 subclass. 
In the spectrum of \#50, the NIII line appears weaker than in \#44 and
for this reason this star is probably cooler than the previous
one. The CIV triplet is still present, but weaker than in \#44,
placing \#50 between the spectral type O6 and O7. 
The combination of features for source \#61 suggest it is the
coolest one in Figure~\ref{spec1}, and we place it in
the kO7--O8 subclass. As in our earlier work, we adopt a ZAMS
classification due to the presence of massive YSOs in the cluster.

\section{W51 IRS 2}

\subsection{High Resolution Images}

W51 IRS 2 (hereafter IRS 2) consists of a number of a stellar objects
surrounded by a peanut--shaped extended emission (as seen in low
resolution images at 2.2 \micron) of $\sim15''\times15''$. Two bright
sources were reported by \citet{gol94} from low resolution
near-infrared images: IRS 2E and IRS2 West (hereafter IRS 2W). IRS2 is
also a radio source, named W51d \citep{martin72}, however the radio
peak does not coincide with the infrared peak \citep{genzel82} IRS 2E.

In this section, we present high resolution images of IRS2 taken with
the adaptive optics near infrared camera NAOS CONICA (NACO)
\citep{lenzen03, rousset03} at the ESO VLT UT4. The raw data were
retrieved from the public database archive (ID 71C-0344(A)). We
processed the data combining the dithered images and subtracting sky
images taken with the same exposure time, but displaced from the
target. The images were flat-fielded and a bad pixel map was applied
to the final combined images. Photometry was performed by means of the
IDL code XSTARFINDER \citep{dio00}, a psf fitting algorithm suitable for adaptive
optics assisted observations. The extracted fluxes were calibrated
using the standard star P9150 \citep{per98} observed as part of the
nightime calibration set. The observing log of the images is presented
in the Table \ref{obslog}. The zoomed box in Figure \ref{color} is a
false color image of IRS2, coded as red ($K_{s}$), green ($H$) and
blue ($J$) and constitutes the highest resolution image obtained from
this region until now 
\citep[the $H-$band image was presented recently by][]{lacy07}.
 
The $H$ and $K$ images of IRS 2 are shown in the
Figure~\ref{nacoh}. In both images, the circles represent the position
of objects detected by \citet{gol94} through a PSF fitting
algorithm. The position of IRS2E is marked by a box and, as can be
seen, three sources lie within it. Goldader \& Wynn-Williams concluded
that IRS2E is a small cluster of stars.  The high resolution images
suggest a somewhat different picture. Three sources are clearly
identified both at $H$ and $K_{S}$. Two of the sources ($a$ and $c$)
appear stellar, while the third ($b$) is an extended clump of
emission.  Furthermore, source $c$ appears relatively blue and may be
a foreground object.  Source $a$ is very red and the brightest $K_{S}$
source in the IRS2 region. We thus associate IRS2E with a single
unresolved source ($a$).

We present the IRS2 sources as open squares on the CMD in the
Figure~\ref{cmd}. We detected 36 sources in all three bands, however 
due to the low quality fitting of the
two brightest sources in the $J$--band, they were removed. 
The source IRS2E, on the other hand is not detected in our 
$J$--band image.

\subsection{Radio Sources and MIR Emission}

IRS 2 hosts three UCHII regions \citep{gaume93}. The high resolution images 
show that a cluster of three objects can be associated with IRS2E within 
1'' (which would be 0.01 pc at a distance of 2.0 kpc). The main radio source
W51d is a cometary UCHII region \citep{wood89}, and the other two
sources are W51d1 and d2 as showed in Figure~\ref{nacoh}. \citet{gol94} 
noted that instead of extended
emission, no NIR counterpart was found at the radio peak position, on
the other hand, \citet{okamoto01} assumed that IRS 2W is the NIR
counterpart of W51d, only by its proximity and by the visual
inspection of the morphology of the extended MIR emission. Figure
\ref{irs2-mir} is a composite image showing both radio and MIR
emission overplotted on the $K_S$--band image of IRS2. The heavy black
contour lines represent MIR emission at arbitrary units, from high
resolution images taken recently at the Gemini South Observatory at
$12.6~\micron$ by us. A MIR study of W51A will be presented in a
forthcoming paper (Barbosa et al. in preparation). The light white
contour lines represent the radio emission, also at arbitrary units,
taken from the radio catalogue of \citet{wood89}. Taking the $K_S$--band
image as reference, we shifted the radio image on 0.8 arcsec to match the 
position of W51d and its NIR counterpart. This value is within the
uncertainties of 1 arcsec as expected by UT4 pointing. 
The MIR image was shifted to match the MIR infrared peak of IRS2E and its NIR 
counterpart. The shift between MIR and NIR images amounts to 2 arcsec. The 
MIR images seem to have an uncertainty worst than the quoted 1 arcsec, 
specially at 25 $\micron$, so we can consider this shift value is within 
the uncertainties. The position of W51d$_{2}$ was derived from its relative 
position to W51d$_{1}$,comparing the coordinates presented by \citet{wood89} 
and \citet{gaume93} for the former source.

The composite image shows IRS2W as the NIR counterpart of W51d, but
also shows that IRS2W is not a MIR source. \citet{okamoto01}, on the
other hand identified IRS2W as the NIR counterpart of their MIR source
OKYM3. The Western MIR emission peaks at the position of two stellar
objects that are also coincident with the secondary peak of radio
emission from W51d. This peak is spatially resolved in the MIR, however, 
and it is not clear if there is a buried source associated with it or if 
it is associated with an ionized cavity produced by IRS2W. We have recently 
obtained a Gemini NIFS spectrum of IRS2W which indicates it is indeed an 
O--type star of the hottest spectral type. This spectrum will be the subject 
of a forthcoming paper.

IRS2E is the brightest source in the $K_S$--band and
at $12.3~\micron$ as well, and it is not a radio source. This lack of
emission at longer wavelengths may be due to its youth, it may be the
case that this object has a low radio luminosity and/or a small
emitting volume giving a low emission measure. However, it is worth
noting that the radio maps presented by \citet{gaume93} show some
extended emission {\it near}, but not coincident, with the position of
IRS2E. Thus we conclude IRS2E is a massive young stellar object
(MYSO), but we can not identify it as an UCHII. W51d$_{1}$ is seen as
a radio, NIR, and MIR source and W51d$_{2}$ was not detected in the
radio maps of \citet{wood89} and may have weak MIR emission.

The high resolution images show that a cluster of three objects can be
associated with IRS2E within 1'', which represents $\sim$0.01 pc at
the new distance of 2.0 kpc. It is also interesting to note that at
the position of the radio source W51d$_{1}$ a small cluster of at
least four sources is seen. A closer inspection of the images reveals
two other sources, but they are too faint to be detected by a PSF
fitting algorithm. We conclude that a small cluster of 4, maybe 6,
sources (all of them within 0.3'' or 660 AU) is the NIR counterpart of
the radio source W51 d$_{1}$. A similar conclusion can be taken from
the near infrared counterpart of source W51 d$_{2}$, since three
sources are seen in its position, all of them separated by less than
0.3''.

\subsection{IRS2E Spectroscopy}

The spectrum of the most luminous source in the $K_{S}$ band in our
sample, IRS 2E, is shown in Figure~\ref{spec2}. The position of this
source in Figure~\ref{cmd} shows that IRS 2E is more luminous in
$K_{S}$ than the O--type stars on the ZAMS which we classified in the
previous section. OB stars can be very luminous in the NIR during
their stage as a MYSO. This high
luminosity is due to the excess emission in the $K$ band
(Figure~\ref{excess}). This $K$-band excess has to be accounted for if
one wants to estimate the spectral type of the MYSO using its
luminosity. The strongest emission lines are Br$\gamma$
($\lambda=$2.166 \micron) and HeI (2.058~\micron~and 2.113~\micron).
Figure \ref{spec2} shows a series of H$_{2}$ and [FeIII] lines. Also,
we confirm the detection of the ``unidentified'' infrared line at
2.288~\micron ~reported by \citet{okumura01}.
This line was discussed earlier by \citet{geballe91} who did not
detect it in IRS2 (apparently due to their large aperture of 5'') but
found it was common in planetary nebulae with middle excitation
temperatures. Based on their observations, \citet{geballe91} were
able to estimate the ionization potential of the species giving rise
to the line which led \citet{din01} to identify it with [Se IV]. The
line is now known to be common in the spectra of planetary nebulae
(PNe) \citep{ster07}. A second line was identified by \citet{din01}
near 2.199~\micron ~as being associated with [Kr III]. This line can be
confused with an H$_2$ transition in lower resolution spectra. Our
spectrum of IRS2E shows a line at this position and it is closer to the
[Kr III] position than the H$_2$ position. Independently of the present
work, \citet{blum07} identified the [Se IV] (and [Kr III]) line in two
UCHII regions. Thus the ``unidentified'' line(s) appears more common
than previously thought and may be produced in higher excitation
regions very close to newly formed massive stars.

The spectrum of IRS 2E is very similar to the spectrum of the UCHII
region G25.2-1.74 presented by \citet{bik05}. Given the lack of a
readily identifiable radio counterpart, this may simply reflect the
nebular nature of IRS~2E and the environment within which it is
found. Disentangling the internal and external excitation of IRS2E is
beyond the scope of this paper.

\section{The Distance to W51A}

In the previous section we classified the spectra of five objects in
W51A: four O--type stars and a MYSO. In this section, the distance to
W51A is determined by using the spectroscopic and photometric results
found for the four O--type stars. Table~\ref{dist} summarizes the main
properties of the O--type stars. We compute distances assuming the
O--type stars shown in Figure~\ref{spec1} are zero--age main--sequence
($D_{ZAMS}$ - column 5 of Table~\ref{dist}) or in the dwarf luminosity
class (column 6). For the dwarf case, the distance is determined using
the $M_V$ given by \citet{mar05} and $V-K$ from \citet{koo83}. 
For the ZAMS case, we use the $M_K$ from \citet{blum00} and 
adjust the $M_K$ to account for the difference in $M_V$ between the 
\citet{vac96} and \citet{mar05} O star properties. \citet{mar05} have 
revised the dwarf O star $M_V$, and we have assumed the ZAMS $M_V$ will 
be similarly shifted. For the spectral types considered here, the 
maximum change is to make the $M_V$ 0.25 mag fainter.

The distance estimates are shown in Table~\ref{dist}. For the derived 
spectral types, we obtain distances of $2.0 \pm 0.3$ and $2.4 \pm 0.4$ kpc 
for the ZAMS and dwarf cases, respectively. The lack of luminosity class
indicators do not allow us to unambigously determine the distance, 
but the ZAMS value is preferred given the presence of massive YSOs in the 
cluster. The uncertainty quoted is the error in the mean.

An extra uncertainty on the derived distance comes from the 
interstellar extinction law. In this paper, we adopted the interstellar 
reddening curve of \citet{mat90}. However, \citet{mes05} and \citet{nis06} 
predict 
$A_{K_{S}}/E_{H-K{S}} \sim 1.5$ in contrast with the value 1.7 suggested by 
\citet{mat90}. Higher values of interstellar extinction 
will result
in a underestimate of the distance. The distances quoted above would be a 
factor of 1.28 larger for the extinction law predictiong a smaller $A_K$.

\subsection{The Lyman Continuum Flux}

\citet{cc04} determined the number of Lyman continuum photons
emitted per second by W51A as 8.71 $\times$ 10$^{50}$ with 
the cluster located at a distance of 5.5 kpc. This corresponds to 
$\approx$ 20 O3V-type stars or $\approx$ 200 O7V stars \citep{mar05}. 
The new spectrophotometric distance we have determined by
using four O--type stars is D = 2.0 $\pm$ 0.3 kpc, considerably
smaller than 5.5 kpc. The new distance brings the number
of Lyman continuum photons down to $NLyC = 1.15 \times 10^{50}$.

The Lyman continuum flux comes primarily from the four brightest stars
which are well above our photometric completeness limit. We have
calculated that the number of Lyman continuum photons from the
contribution of all massive stars we have identified in the cluster is
$NLyC = 1.5 \times 10^{50}$ photons s$^{-1}$.

The observed NLyc photons derived from radio techniques is only a
lower limit to that emitted by the stars, since some may be destroyed
by dust grains or leaked through directions of low optical depth. On
the other hand, due to the high extinction and crowding in the region
some ionising stars may not have been detected in the $J$, $H$ and
$K$--bands. Therefore, both values found for the total $NLyC$ produced
by the hot stars in W51A, by radio techniques and near--infrared,
should be considered as lower limits. The $NLyC$ photons from the
radio measurements and that derived from the star counts are in good
agreement.

\section{Discussion and Summary}

We have presented $J$, $H$ and $K_S$ images of the HII region W51A
obtained on the 4-m Blanco Telescope using the facility imager ISPI.
The photometric results agree with the previous work, such as
\citet{okum00}. $K$--band spectra of five of the brightest cluster
members were obtained with Gemini North telescope using the instrument
NIRI.  Four of them have been classified as O--type stars and one as a
massive young stellar object. We have derived the distance to W51A by
using the spectroscopic parallaxes of four O--type stars. We also have
presented archival images of the compact HII region IRS 2 taken with
the adaptive optics camera NACO at the VLT. The high resolution images
revealed a small cluster of four objects as the NIR counterpart of the
UCHII region W51d$_{1}$ and another cluster with three objects at the
position of the UCHII region W51d$_{2}$. The source IRS 2E was
resolved into three sources and we identified the brightest source in
the $K$ band as IRS 2E. The $K$ band spectrum of IRS 2E is very
similar to that of an UCHII region, however, we do not explicitly
classify it as so, because there is no distinct radio source with it.

The value we have derived for the distance to W51A by spectroscopic
parallaxes of O--type stars, 2.0 $\pm$ 0.4 kpc, is considerably
smaller than the kinematic distance: 5.5 kpc \citep{cc04,russ03}. 
The systematic uncertainties on the interstellar extinction law are not big
enough to change this picture. As a
consequence of this result, W51A is fainter than previously
determined. Although the number of Lyman continuum photons is smaller
with the new distance, $\sim$ 1.5~$\times$~10$^{50}$ photons per second,
W51A can still be classified as a giant HII region harboring a
considerable number of massive stars.

The newly derived distance is much smaller than the distance (7 $\pm$
1.5 kpc) derived from H$_2$O maser observations presented by
\citet{genzel81,genzel82} for the masers associated with W51A MAIN (about 1$'$
south east of IRS2E) and the similar result (8.5 $\pm$ 2.5 kpc) for
W51A NORTH derived by \citet{schneps81}. The method used by
\citet{genzel82} and \citet{schneps81} assumes the maser motions are random
and isotropic. \citet{imai02} present somewhat more detailed models of
the maser kinematics in W51A NORTH (centered within 1$''$ of IRS2E)
and find that the motions are better explained by a radial out flow
model which they use to fit the positions and outflow velocity. 
The distance derived from this technique is 6 $\pm$ 1.3 kpc, very close to
the kinematic distance. On the other hand, they suggest the distance is 
completely covariant with the expansion velocity and the direction of the 
flow and the uncertainty permits a distance value of 7 kpc.
\citet{imai02} further show that the masers
for W51A MAIN and NORTH are non--spherically symmetric and that there
are significant residuals with respect to the assumption that all the
maser spots emminate from a single point at the origin of the out flow
(i.e. there are signiticant deviations for a simple radial out ward
flow when looking at the derived maser space positions and velocity
vectors). Using this assumption of random and isotropic motion for 
the water maser, \citet{imai02} estimated the distance of 4 kpc to 
W51A NORTH. This is in disagreement with the value found previously 
by \citet{genzel81}. It is not clear what other models could fit the 
data or how serious the non--symmetry of the flow is in comparison 
to the model assumptions. In particular, it is not clear if a near 
distance model could fit the data. 

The situation may be even more complex in light of recent mid infrared
observations (in [Ne II]) made by \citet{lacy07}. These investigators
discovered a large ``fan like jet'' or outflow which is ionized as it
enters the HII region (apparenly from a source embedded in the
molecular cloud) very close to W51A NORTH. They speculate that the
ionized flow may have the same source as the W51A NORTH masers and
this would require precession of the outflow since the ionized flow
has a strong radial velocity while the masers are dominated by proper
motions in the plane of the sky.

\begin{figure}[find] \begin{center}
\includegraphics[totalheight=10.0cm,angle=0]{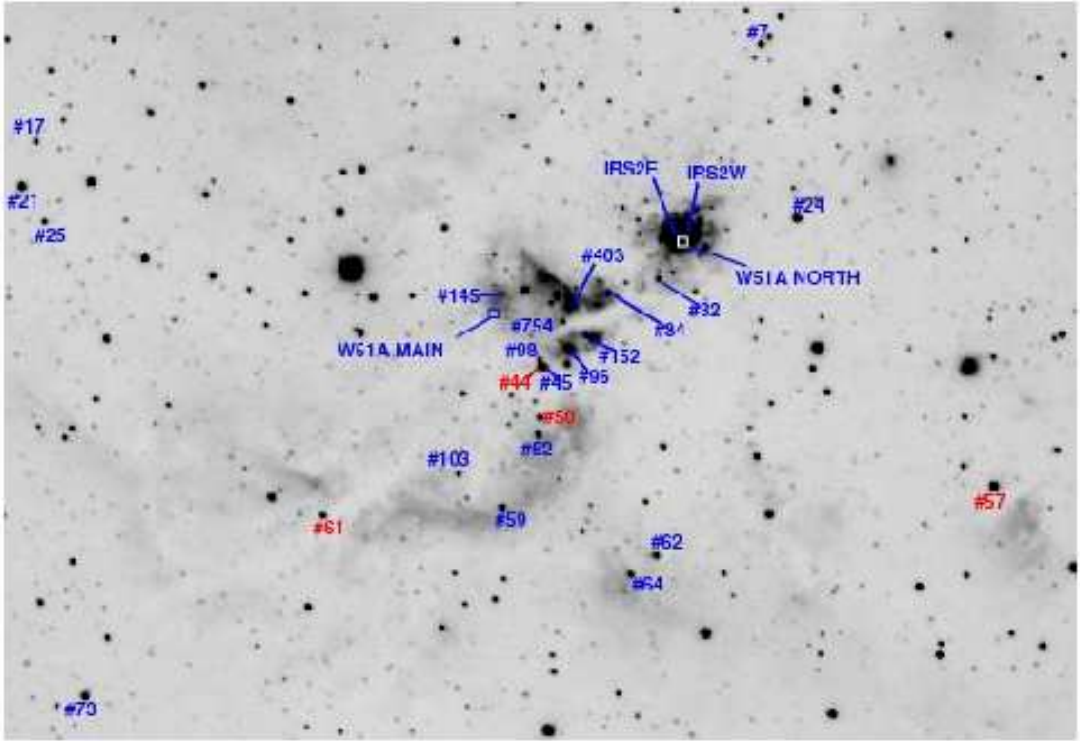} \end{center}
\caption{Finding chart using a $K_S$--band image of W51A, where DoPHOT
photometry was performed (red rectangle in Figure~\ref{color}). Object
labels refer to the brightest cluster members detected by DoPHOT.  
{\it Squares} refer to the masers W51A NORTH and MAIN (see \S7).
stars \#44, \#50, \#57 and \#61 whose spectra show photospheric lines
typical of O--type stars, have been used to determine the
spectroscopic parallax of W51A.  The size of the image corresponds to
an area of $\approx 5.1' \times 3.5'$.  North is up and East to the
left. \label{finding}}
\end{figure}

\begin{figure}[cmd] \begin{center}
\includegraphics[totalheight=15.0cm,angle=-90]{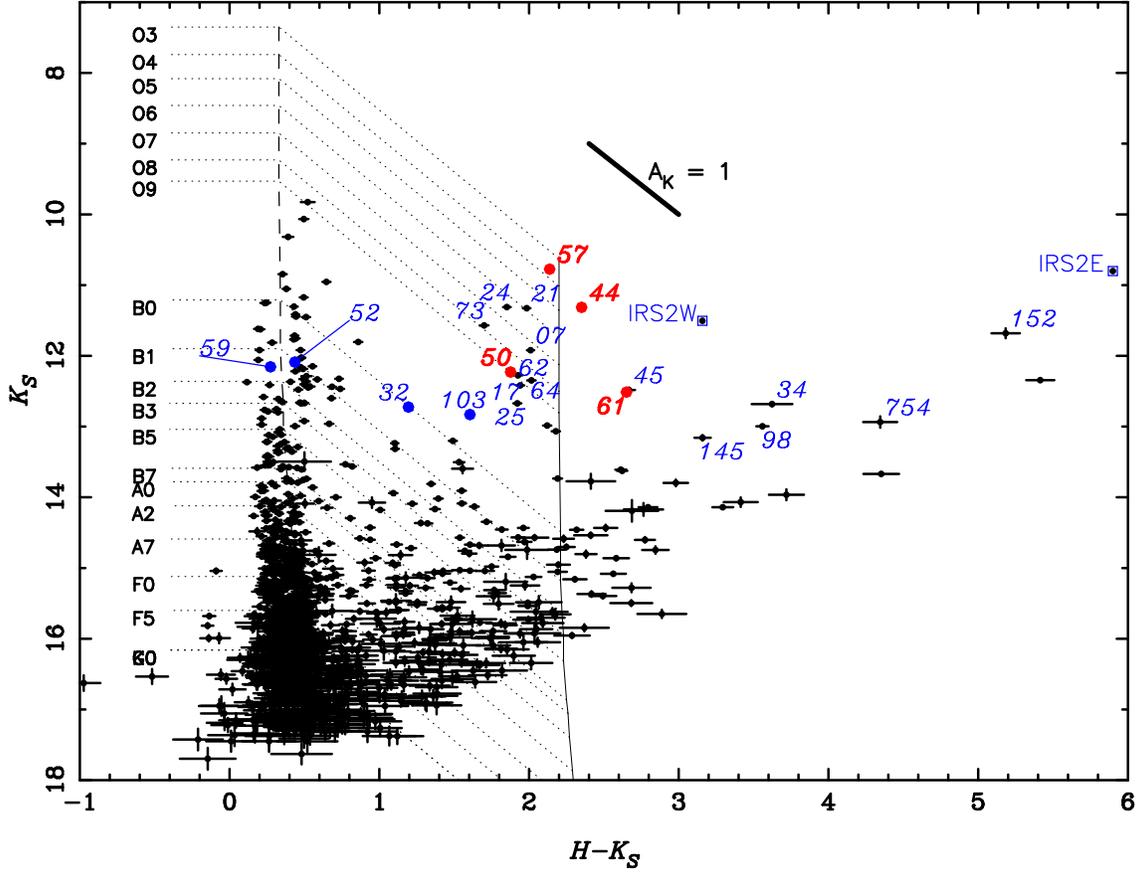}
\end{center} \caption{$K_S$ vs $H - K_S$ color--magnitude diagram (CMD) of W51A
showing the position of the theoretical ZAMS shifted to 2.0 kpc and
with interstellar extinction $A_{K_S} = 0.65~mag$ (dashed line). An
additional ``cluster'' extinction component of $A_{K_S} = 3.27~mag$
($A_{K_S total} = 3.92~mag$) results in the ZAMS position indicated by
the vertical {\it solid} line. Object number labels are the same as in
Figure~\ref{finding}. {\it Open
squares} indicate sources detected in IRS2 (see zoomed box in
Figure~\ref{color} and \S5.1 for more details).\label{cmd}} 
\end{figure}

\begin{figure}[ccd] \begin{center}
\includegraphics[totalheight=15.0cm,angle=-90]{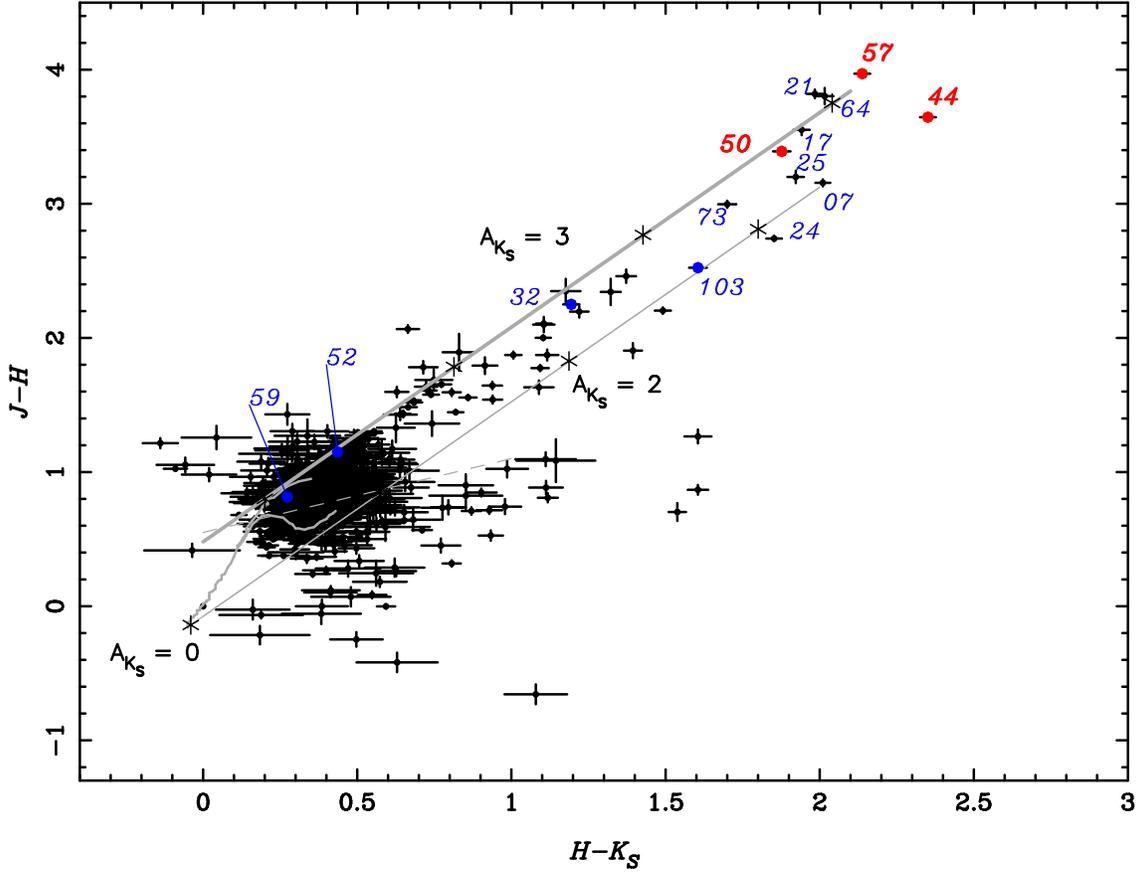}
\end{center} \caption{$J-H$ vs $H-K_S$ color-color plot showing the reddening
line of M--type stars ({\it heavy solid} line), O--type stars ({\it
solid} line) and T~Tauri stars ({\it dashed} line). {\it Dots} refer
to stars detected by DoPHOT in all three bands $J$, $H$ and
$K_S$. The {\it asterisks} indicate the
corresponding $A_{K_{S}}$ along the reddening vector.  Object number
labels are the same as in Figure~\ref{finding}
and~\ref{color}. \label{ccd}}
\end{figure}

\begin{figure}[ccd] \begin{center}
\includegraphics[totalheight=15.0cm,angle=-90]{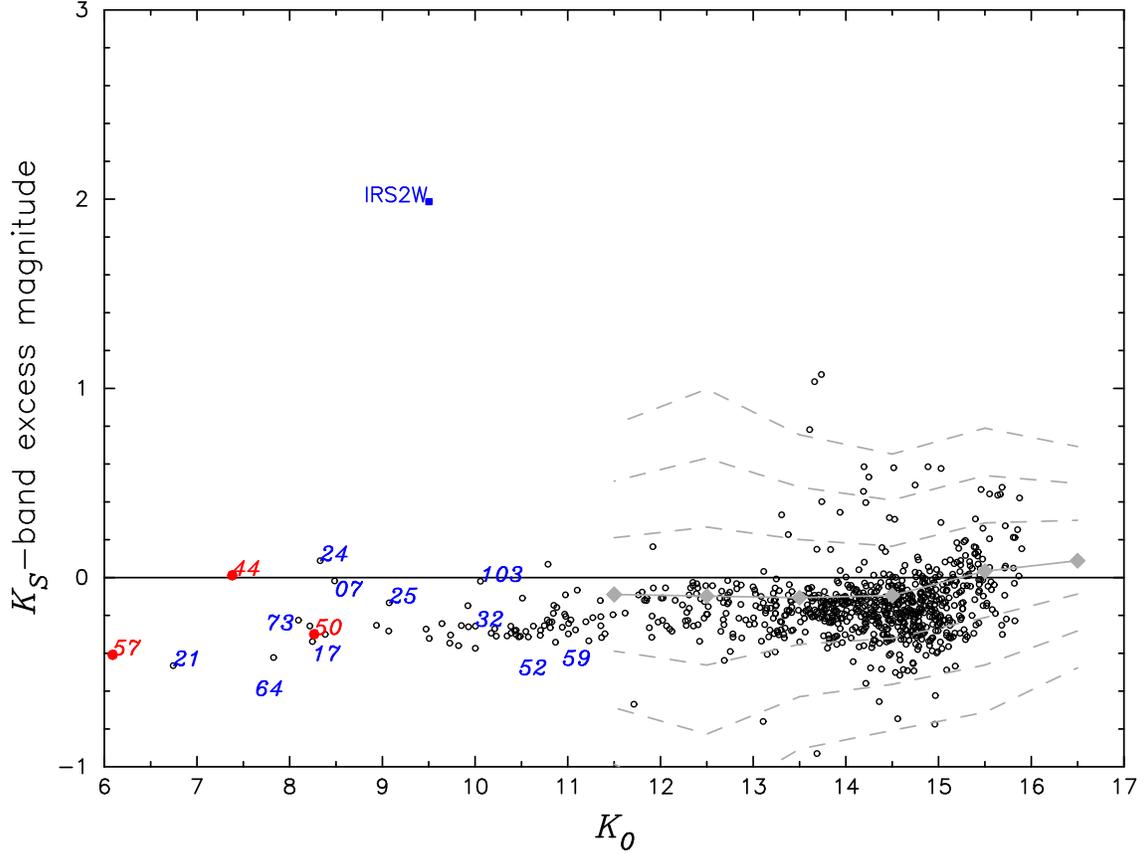}
\end{center} \caption{Excess emission as a function of dereddened $K_S$--band
magnitude ($K_{\circ}$). Only objects detected in all three bands have been 
included. Circles indicate objects with measured $J$,
$H$ and $K_S$ magnitudes. The solid line indicates
$K_{exc}=0$. Connected solid diamonds refer to the average value of
the $K_S$ excess in one magnitude bins. A 2 sigma clipping was applied 
to calculate the average valuesDashed lines indicate one,
two, and three $\sigma$ deviations from the average. IRS2W ({\it
Solid squares}) have been detected in the VLT images (See \S5.1 for
more details). Very positive values represent circumstellar excess
emission while small or slightly negative values do not; see text.
\label{excess}} 
\end{figure}

\begin{figure}[!ht]
\centerline{\includegraphics[totalheight=18.0cm,angle=0]{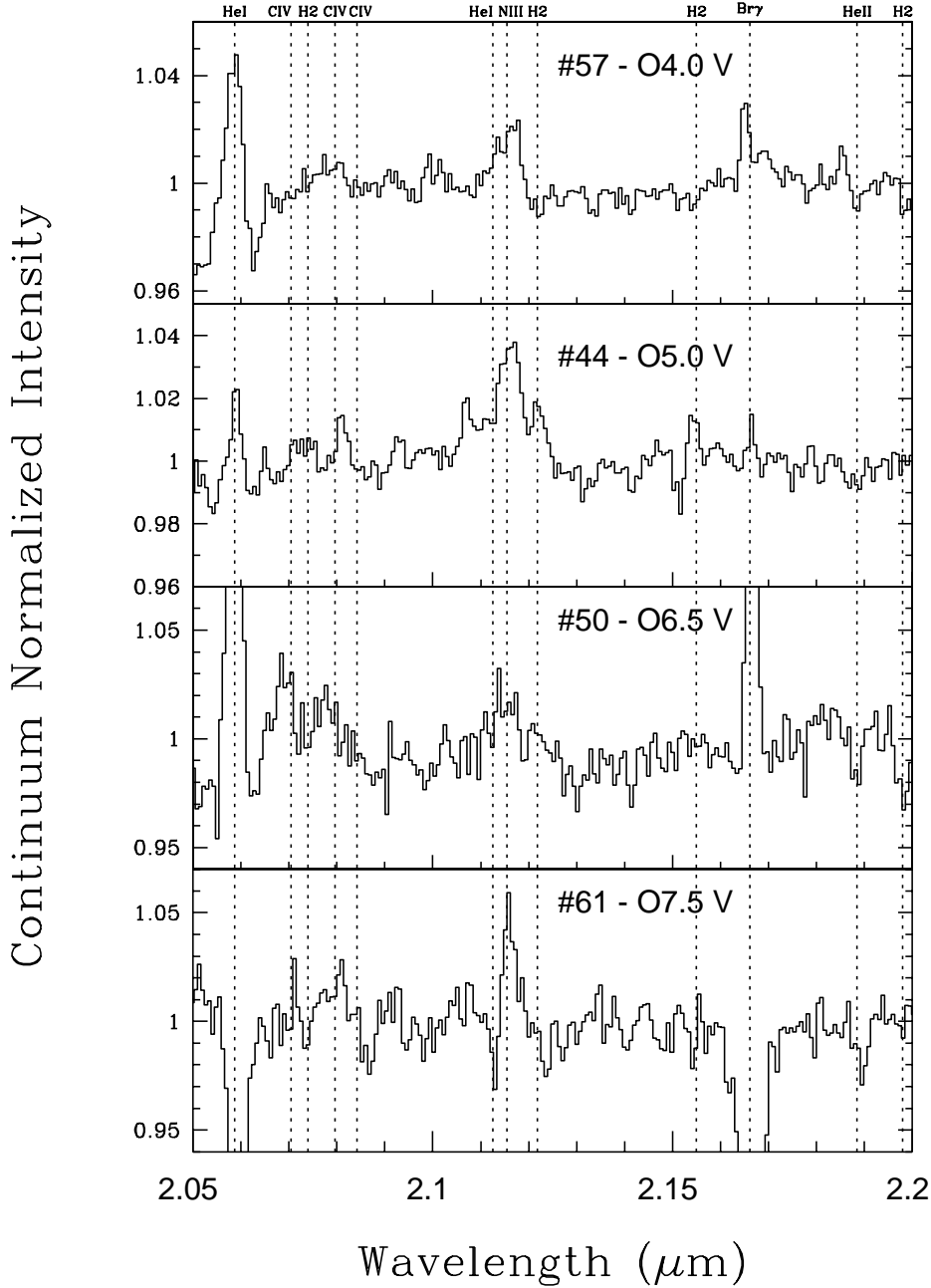}}
\caption{$K$--band spectra for the four brightest stars in the W51A:
\#57 (O4V), \#44 (O5V), \#50 (O6.5V) and \#61 (O7.5V). The spectral
type was determined by comparison with the O--type stars from
\citet{han96}. The spectral resolution of $R$ $\approx$ 780
gives a linear dispersion of $\approx$ 7.08 \AA /pixel. Spectra were
summed in apertures $0''.96$ wide (eight pixels) by a slit width of
$0''.48 \times 110''$ and include background subtraction from
apertures centered $\le 1''.0$ on either side of the object.  Each
spectrum has been normalized by a low--order fit to the continuum
(after correction for telluric absorption). The spectra are often
contaminated by the 2.058 $\mu$m HeI and Br$\gamma$ nebular lines
which can be over or under subtracted from the stellar source. The
signal to noise ratio is $S/N \sim 100-120$.\label{spec1}}
\end{figure}

\begin{figure}[!ht]
\centerline{\includegraphics[height=16.0cm,width=7.0cm,angle=-90]{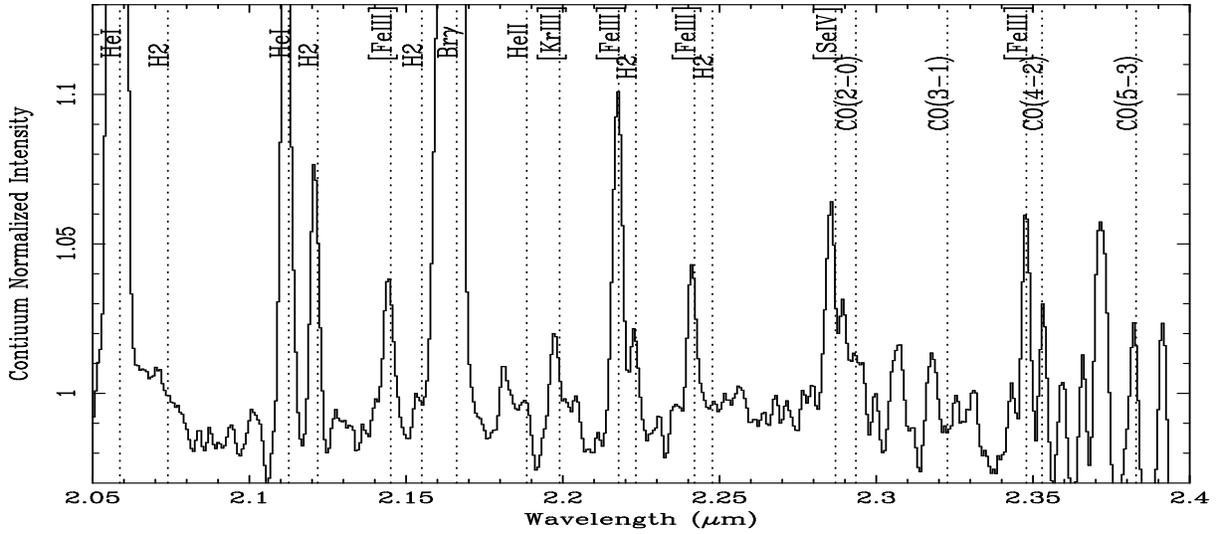}}
\caption{$K$--band spectrum of IRS2E taken with the same setup as the
O--type star spectra shown in Figure~\ref{spec1}. \label{spec2}}
\end{figure}

\begin{figure}[!ht]
\centerline{\includegraphics[totalheight=9.0cm,angle=-90]{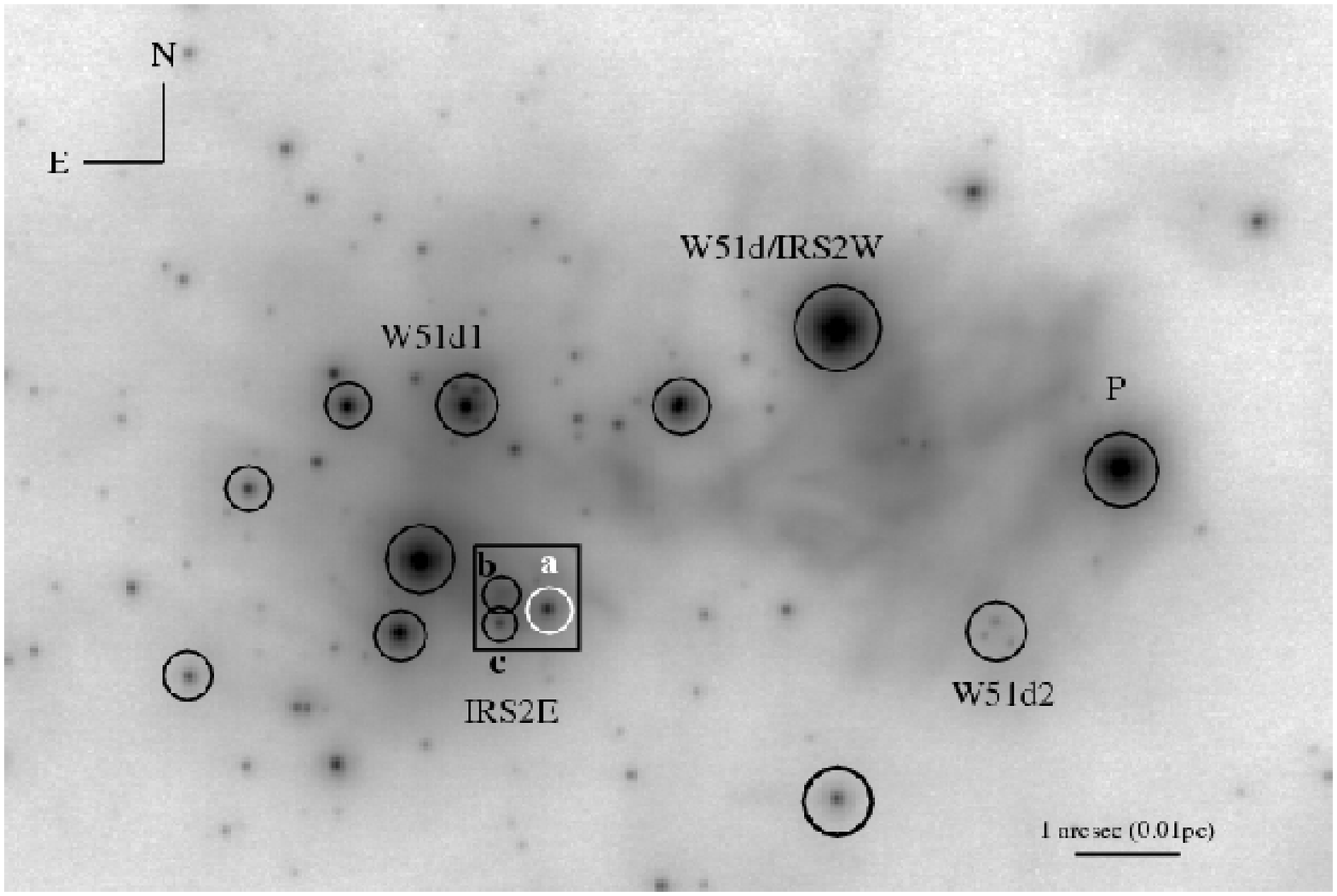}
\includegraphics[totalheight=9.0cm,angle=-90]{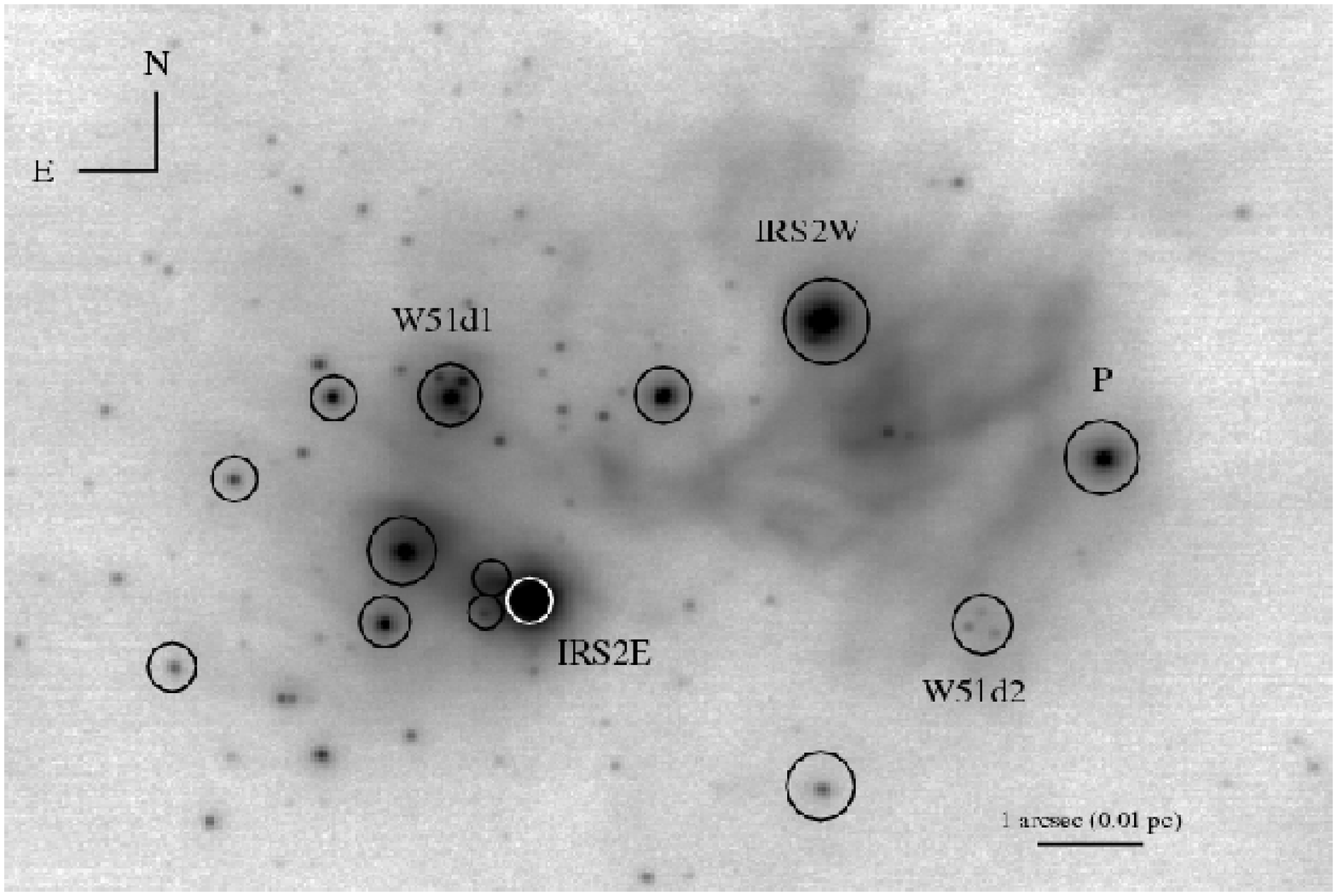}}
\caption{W51 IRS2 in the $H$--band (left pannel) and $K_{S}$--band
(right pannel). The circles labeled a, b and c refer to the position of 
IRS2E, an extended emission (no stellar source) and a NIR source also detected by 
\citet{gol94} respectively. The star labeled ``P'' is also detected on the POSS red
plate.  The position of IRS2E is marked by a box (sized
1''$\times$1''). IRS2E is the brightest source in the field. The NIR
counterparts of radio sources W51d, d$_{1}$ and d$_{2}$ are also
indicated.
\label{nacoh}}
\end{figure}

\begin{figure}[!ht]
\centerline{\includegraphics[totalheight=8.0cm,angle=0]{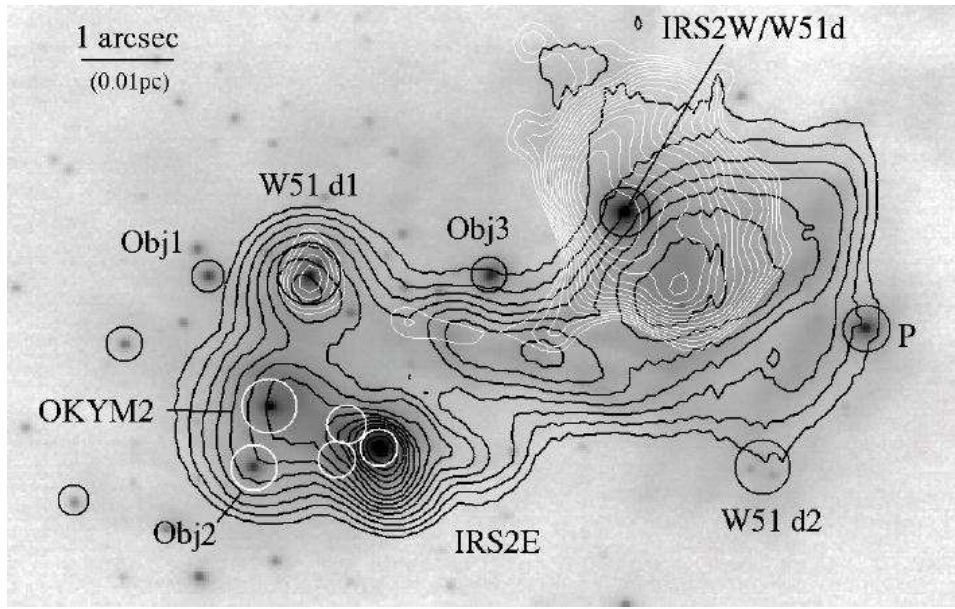}}
\caption{Composite image of IRS 2E. The $K_{S}$--band image is
overplotted with radio emission contours from \citet{wood89} in
arbitrary units in light white contours from \citet{wood89} and by MIR
emission contours (C. Barbosa, unpublished), also at arbitrary units
in heavy black contours. Sources are indicated by black or white circles 
to give the best contrast. W51d2 and the radio emission southern of
W51d1 were not detected by \citet{wood89}, but fainter radio emission
does extend toward IRS2E in the maps of \citet{gaume93}.
\label{irs2-mir}}
\end{figure}

The masers of W51A NORTH are probably associated with IRS~2, while the
O stars are somewhat further away in projection from IRS~2. The masers
and O stars may trace different physical locations along the spiral
arm and appear close only in projection.  Within the uncertainty of
the maser techniques, this explanation is less favorable since the
distance derived for W51A MAIN is similar to the larger distance to
W51A NORTH and W51A MAIN is closer (in projection) to several of the O
stars (see Figure~\ref{finding}).
If the O stars are actually unresolved multiple (massive) stars they
would appear brighter and thus apparently closer. On average, this
would require that nine stars of similar luminoisty make up each
object we have observed if we consider just the ratio of distances two
kpc to six kpc. This seems an unlikely explanation.  We are presently
left with no satisfactory explanation for the distance discrepancy
between the masers and radio observations on the one hand, and the
near infrared spectrophotometric observations on the other.

The authors wish to thank Ed Churchwell for making available the radio
map of W51 IRS2 in electronic form. We thank an
anonymous referee for the careful reading of this paper and for the
useful comments and suggestions which have resulted in a much improved
version. AD and CLB are grateful 
to FAPESP and CNPq for support. PSC appreciates continuing support from the
NSF. These results are based on observations obtained at the Gemini
Observatory. The Gemini Observatory is operated by the Association of
Universities for Research in Astronomy, Inc., under a cooperative
agreement with the NSF on behalf of the Gemini partnership: the
National Science Foundation (United States), the Science and
Technology Facilities Council (United Kingdom), the National Research
Council (Canada), CONICYT (Chile), the Australian Research Council
(Australia), CNPq (Brazil), and CONICET (Argentina). This paper is
partially based on observations made with ESO Telescopes at the
Paranal Observatories under programme 71C-0344(A).

\clearpage

\clearpage

\pagestyle{empty} 
\begin{deluxetable}{lrcrcccc} 
\tablecaption{Colors and Magnitudes\tablenotemark{*} ~Catalog 
\label{stars}}  
\tablewidth{0pt}
 \tablehead{
\colhead{ID} & 
\colhead{$J$} & 
\colhead{$H$} & 
\colhead{$K_S$} & 
\colhead{$J-H$} & 
\colhead{$H-K_S$}} 
\startdata 
IRS2E\tablenotemark{**} & $>$ 21 & 16.70 $\pm 0.01$ &  10.80 $\pm 0.01$ & $<$ 4.3  & 5.90 $\pm 0.01$ \\ 
IRS2W\tablenotemark{**} & 16.47 $\pm 0.01$ & 14.66 $\pm 0.01$ & 11.50 $\pm 0.01$ & 1.81 $\pm 0.01$  &  3.16 $\pm 0.01$\\ 
\#7   & 17.09 $\pm$ 0.02 & 13.93 $\pm$ 0.01 & 11.92 $\pm$ 0.02 & 3.16 &  2.01 \\ 
\#17  & 17.91 $\pm$ 0.04 & 14.36 $\pm$ 0.02 & 12.42 $\pm$ 0.02 & 3.55  &  1.94 \\ 
\#21  & 17.13 $\pm$ 0.03  & 13.31 $\pm$ 0.01 & 11.33 $\pm$ 0.02 & 3.82  &  1.98 \\ 
\#24  &  15.90 $\pm$ 0.01 & 13.16 $\pm$ 0.01 & 11.31 $\pm$ 0.02 & 2.74  & 1.85 \\ 
\#25  & 17.79 $\pm$ 0.04 & 14.59 $\pm$ 0.02 & 12.67 $\pm$ 0.02 & 3.20  &  1.92 \\ 
\#32  & 16.17 $\pm$ 0.01 & 13.92 $\pm$ 0.01 & 12.73 $\pm$ 0.02 & 2.25  &  1.19 \\ 
\#34  & \nodata          & 16.31 $\pm$ 0.13 & 12.68 $\pm$ 0.03 & \nodata &  3.62  \\ 
\#44  & 17.31 $\pm$ 0.01 & 13.67 $\pm$ 0.01 & 11.31 $\pm$  0.02 & 3.65  &  2.35  \\ 
\#45  & \nodata          & 15.15 $\pm$ 0.02  & 12.48 $\pm$ 0.03 & \nodata &  2.67  \\ 
\#50  & 17.50 $\pm$ 0.02 &  14.11 $\pm$ 0.02 & 12.23 $\pm$ 0.02 & 3.39  &  1.88  \\ 
\#52  &  13.67 $\pm$ 0.01 & 12.53 $\pm$ 0.01 & 12.09 $\pm$ 0.02 & 1.15  &   0.44  \\ 
\#57  & 16.88 $\pm$ 0.02 & 12.91 $\pm$ 0.01 & 10.77 $\pm$ 0.02 & 3.97  &  2.14  \\
\#59  & 13.24 $\pm$ 0.01 & 12.43 $\pm$ 0.01  & 12.16 $\pm$ 0.02 & 0.81  &  0.27  \\ 
\#61  & \nodata          &  15.16 $\pm$ 0.02 & 12.51 $\pm$ 0.02 & \nodata &  2.65  \\ 
\#62  & \nodata          & 14.20 $\pm$ 0.02 & 12.28 $\pm$ 0.02 & \nodata &  1.93  \\ 
\#64  & 18.17 $\pm$ 0.06 & 14.36 $\pm$ 0.02 & 12.35 $\pm$ 0.02 & 3.80  &  2.02  \\ 
\#73  & 16.27 $\pm$ 0.02 & 13.27 $\pm$ 0.02  & 11.57 $\pm$ 0.02 & 3.00  &  1.70   \\ 
\#98  & \nodata          & 16.56 $\pm$  0.03 & 13.00 $\pm$ 0.02 & \nodata     &  3.56   \\ 
\#103 & 16.96  $\pm$ 0.01 & 14.44 $\pm$ 0.02 & 12.83 $\pm$ 0.02 & 2.52 &  1.61	\\  
\#145 & \nodata          & 16.32 $\pm$ 0.04 & 13.16 $\pm$ 0.04 & \nodata     &  3.16   \\
\#152 & \nodata          & 16.86 $\pm$ 0.06  & 11.68 $\pm$ 0.07 & \nodata     & 5.18   \\ 
\#754 & \nodata          & 17.28 $\pm$ 0.08 & 12.94 $\pm$ 0.08 & \nodata      &  4.35   \\ 
\enddata  

\tablenotetext{*}{The sum in quadrature of the zero point uncertainty
 plus the PSF--fitting uncertainty define the final magnitude errors
 in $J$, $H$ and $K_S$, see \S2.}
 
\tablenotetext{**}{VLT data}

\end{deluxetable}

\begin{deluxetable}{lcccc} 
\tablecaption{Equivalent width (\AA) of strategic lines used to classify the
O-type stars \label{EWS}}  \tablewidth{0pt} \tablehead{\colhead{Lines} &
\colhead{\#57} & \colhead{\#44} & \colhead{\#50} & \colhead{\#61}}  \startdata
HeI (2.059$\mu$m) & 1.75 $\pm$ 0.13   & 1.42 $\pm$ 0.22  & 5.35 $\pm$ 0.37   &
6.21  $\pm$ 0.28*ab \\ CIV (2.071$\mu$m) & -              & 0.37 $\pm$
0.05  & 0.72 $\pm$ 0.14*b  & 0.24  $\pm$ 0.12 \\ CIV (2.080$\mu$m) & 0.32 $\pm$
0.04*b  & 0.96 $\pm$ 0.08  & 0.53 $\pm$ 0.14   & 0.61  $\pm$ 0.17 \\ NIII
(2.116$\mu$m)& 0.95 $\pm$ 0.05   & 1.82 $\pm$ 0.09  & 0.64 $\pm$ 0.11*b  &
1.70  $\pm$ 0.21 \\ Brg (2.166$\mu$m) & 0.68 $\pm$ 0.09   & 0.41 $\pm$ 0.09  &
3.83 $\pm$ 0.22   & 11.23 $\pm$ 0.20*ab \\ HeII (2.189$\mu$m)& 0.21 $\pm$
0.10*ab & -      & 0.51 $\pm$ 0.16*ab & 0.63  $\pm$ 0.12*ab \\
\enddata  \tablenotetext{*}{b -- blended lines} \tablenotetext{*}{ab --
absortion lines} \end{deluxetable}

\begin{deluxetable}{lccccccc}
\tablecaption{NACO observations of W51 IRS2: instrumental set-up and observation
logs. Camera S27 provides images with resolution of 27.15 mas/pixel and FOV of
$28 \times 28$ arcsec employing an Aladdin InSb 1024 $\times$ 1024 pixel array.
\label{obslog}}
\tablewidth{0pt}
\tablehead{
\colhead{Filter} & \colhead{Central Wavelength} & \colhead{Bandwidth} & 
\colhead{DIT}\tablenotemark{a} & \colhead{NDIT}\tablenotemark{b} & 
\colhead{PSF}\tablenotemark{c} & \colhead{Date of} & 
\colhead{Seeing}\tablenotemark{d} \\ 
\colhead{} & \colhead{($\micron$)} & 
\colhead{($\micron$)} & \colhead{(s)} & \colhead{(s)} & \colhead{(arcsec)} & 
\colhead{observation} & \colhead{(arcsec)}}
\startdata 
$J$ & 1.27 & 0.25 & 20 & 2 & 0.17 & 07/22/03 & 0.41 \\ 
$H$ & 1.66 & 0.33 & 20 & 2 & 0.09 & 07/21/03 & 0.36 \\ 
$K_{s}$ & 2.18 & 0.35 & 20 & 2 & 0.08 & 07/22/03 & 0.36 \\ 
\enddata
\tablenotetext{a}{DIT corresponds to the exposure time of one image.}
\tablenotetext{b}{NDIT corresponds to the number of exposures.}
\tablenotetext{c}{PSF corresponds to the average PSF of detected objects in the field.}
\tablenotetext{d}{Seeing measured by the ESO Health Check Monitor at each band.}
\end{deluxetable}

\begin{deluxetable}{lcccccccc} 
\tablecaption{Properties of O-type stars in W51A\label{dist}}  
\tablewidth{0pt} 
\tablehead{ \colhead{ID} & 
\colhead{$K_S$}& 
\colhead{$H-K_S$} & 
\colhead{$A_{K_S}$\tablenotemark{a}} &  
\colhead{$D_{ZAMS}$ (kpc)} &
\colhead{$D_V$ (kpc)} & 
\colhead{Spectral Type}} 
\startdata 
\#44  & 11.31 & 2.35 & 4.08 & 1.7 & 2.0 & O5 \\ 
\#50  & 12.23 & 1.88 & 3.28 & 2.8 & 3.6 & O6.5 \\ 
\#57  & 10.77 & 2.14 & 3.72 & 1.9 & 2.1 & O4 \\ 
\#61  & 12.51 & 2.65 & 4.59 & 1.4 & 2.0 & O7.5 \\ 
\\ 
Average & & 2.25 & 3.92 & 2.0 $\pm$ 0.3 \tablenotemark{b}  & 2.4 $\pm$ 0.4
\tablenotemark{b} &   
\enddata  
\tablenotetext{a}{$A_{K_S}$ was determined from $H-K_S$, using $(H-K_S)_0 =
-0.05$ and the extinction law of \citet{mat90}}
\tablenotetext{b}{The uncertainty quoted is the error in the mean.}
\end{deluxetable}

\end{document}